\documentclass[12pt]{article}
\usepackage{amssymb}
\usepackage{amsmath}
\def\be{\begin{equation}}
\def\ee{\end{equation}}
\def\arr{\begin{array}{rll}}
\def\ea{\end{array}}
\def\bea{\begin{eqnarray}}
\def\eea{\end{eqnarray}}

\def\N2{$N{=}2$}

\def\>{\rangle}
\def\<{\langle}
\def\+{\dagger}
\def\={\ =\ }

\begin{document}
\renewcommand{\thefootnote}{\arabic{footnote}}
\begin{titlepage}
\setcounter{page}{0}
\begin{flushright}
$\quad$  \\
\end{flushright}
\vskip 3cm
\begin{center}
{\LARGE\bf Superfield approach to higher derivative}\\
\vskip 0.5cm
{\LARGE\bf $\mathcal{N}=1$ superconformal mechanics}\\
\vskip 2cm
$
\textrm{\Large Ivan Masterov and Boris Merzlikin}
$
\vskip 0.5cm
{\it
Tomsk State University of Control Systems and Radioelectronics,\\
634050, Tomsk, Lenin Ave. 40, Russia}
\vskip 1cm
{E-mails: iv.masterov@yandex.ru, merzlikin@tspu.edu.ru}

\end{center}
\vskip 1cm
\begin{abstract}
\noindent
We formulate the equations which determine a potential function in an $\mathcal{N}=1$ higher derivative supersymmetric mechanics 
compatible with the $osp(2|1)\oplus so(d)$ symmetry and provide a few explicit examples.
\end{abstract}

\vskip 1cm
\noindent
PACS numbers: 02.20.Sv, 11.30.-j, 11.30.Pb

\vskip 0.5cm

\noindent
Keywords: superconformal mechanics, $l$-conformal Galilei algebra, supersymmetry

\end{titlepage}
\noindent
{\bf 1. Introduction}
\vskip 0.5cm
\noindent
Superconformal mechanical systems have attracted considerable attention over the last three decades. The very first examples in the literature provided
supersymmetric counterparts for the $d=1$ de Alfaro-Fubini-Furlan conformal mechanics
\cite{Akulov_Pashnev}-\cite{Ivanov_Krivonos_Leviant} and the Calogero model \cite{Freedman_Mende}. Recent studies of the nonrelativistic
version of the AdS/CFT-correspondence\footnote{See, e.g., Ref. \cite{Dobrev} and references therein.} and a possible link to the black holes physics \cite{Claus_Derix}-\cite{Maloney_Spradlin} have brought the $d=1$ superconformal
mechanics into focus again.\footnote{There exists a rather extensive
literature on the subject. For a review and further references see, e.g., Refs. \cite{Galajinsky_1}-\cite{Galajinsky_Lechtenfeld_2}.}

In parallel, there has been considerable progress in exploration of superconformal mechanics in $d>1$
\cite{Gomis_Townsend}-\cite{Fedoruk_Ivanov_2}. In this context, supersymmetric extensions of the so-called $l$-conformal Galilei algebra
\cite{Henkel}-\cite{Negro_1} and its dynamical realizations play the key role.

The $l$-conformal Galilei algebra, where $l$ is a positive (half)integer parameter, represents a finite-dimensional
conformal extension of the Galilei algebra \cite{Negro_1}. It includes
\bea\label{generators}
\begin{aligned}
&
\bullet && \mbox{the generator of time translations} && L_{-1},
\\[2pt]
&
\bullet && \mbox{the generator of dilatations} && L_{0},
\\[2pt]
&
\bullet && \mbox{the generator of special conformal transformations} && L_{1},
\\[2pt]
&
\bullet && \mbox{the chain of vector generators }(n = 0,1,..,2l) && C_i^{(n)},
\\[2pt]
&
\bullet && \mbox{the generators of spatial rotations} && M_{ij},
\end{aligned}
\eea
where $i,j=1,\dots,d$ and $d$ is the spatial dimension.
The algebra is not semisimple. Its semisimple part is given by $so(1,2)\oplus so(d)$ while the abelian ideal is formed by the vector generators
$C_i^{(n)}$. Subalgebras $so(1,2)$ and $so(d)$ are spanned by the generators $L_n$ $(n=-1,0,1)$ and $M_{ij}$, respectively.

The case of $l=\frac{1}{2}$ is known as the Schr\"{o}dinger algebra, as it comprises
the symmetry of the Schr\"{o}dinger equation associated with a free massive particle \cite{Niederer}. The instance of $l=1$ is conventionally referred to as the conformal Galilei algebra. It is regarded as the nonrelativistic counterpart of the relativistic conformal algebra $so(d+1,2)$
\cite{Lukierski_Stichel}.

Recently, there has been an upsurge of interest in dynamical realizations of the $l$-conformal Galilei algebra for $l>1$
\cite{Fedoruk_Ivanov_3}-\cite{Andrzejewski_1}. In general, the order of differential equations which govern such a system correlates with the value of $l$.
In particular, it was argued in \cite{Gomis_Kamimura} that a free higher derivative particle described by the action functional\footnote{Throughout the work we use the notation $f^{(n)} = \frac{d^n f}{dt^n}$. Summation over repeated spatial indices is understood.}
\bea\label{bosonic_action}
S=\frac{1}{2}\int dt\,\lambda_{ij} x_i x_j^{(2l+1)},\quad \lambda_{ij}=\left\{
\begin{aligned}
&
\delta_{ij} &&i,j=1,2..,d, && \mbox{for half-integer $l$},
\\[2pt]
&
\epsilon_{ij} && i,j=1,2, && \mbox{for integer $l$},
\end{aligned}\right.
\eea
possesses the $l$-conformal Galilei symmetry. In (\ref{bosonic_action}) $\epsilon_{ij}$ is the Levi-Civit\'{a} symbol with $\epsilon_{12}=1$.

$\mathcal{N}=1$, $\mathcal{N}=2$, and $\mathcal{N}=4$ supersymmetric extensions of the $l$-conformal Galilei algebra for $l>1$ have been constructed in Refs. \cite{Masterov_5}-\cite{Aizawa_3}.
In contrast to $\mathcal{N}=2$ and $\mathcal{N}=4$ cases, there is a unique $\mathcal{N}=1$ superextension.
Apart from the generators (\ref{generators}), $\mathcal{N}=1$ $l$-conformal Galilei superalgebra includes
\bea\label{odd_generators}
\begin{aligned}
&
\bullet && \mbox{the generator of supersymmetry transformations} && Q_{-1/2},
\\[2pt]
&
\bullet && \mbox{the generator of superconformal transformations} && Q_{1/2},
\\[2pt]
&
\bullet && \mbox{the chain of odd vector generators }(n = 0,1,..,2l-1) && L_i^{(n)}.
\end{aligned}
\nonumber
\eea
$Q_{-1/2}$, $Q_{1/2}$ together with $L_{-1}$, $L_{0}$, and $L_{1}$ form $osp(2|1)$ subalgebra while $C_i^{(n)}$ and $L_{i}^{(n)}$ enter
abelian ideal of the superalgebra.

In Ref. \cite{Masterov_1}, $\mathcal{N}=1$, $l > 1/2$ conformal Galilei superalgebra  was identified with the symmetry algebra of a free $\mathcal{N}=1$ higher derivative superparticle.
The goal of the present paper is to construct a superfield formulation for that model and to describe potential functions which preserve $osp(2|1)\oplus so(d)$ symmetry. 
It is presented in the next section. Some explicit examples of the $osp(2|1)\oplus so(d)$-invariant models are given in Sect.~3. In the concluding Sect.~4, we summarize our results and discuss further possible developments. Some technical details are given in Appendix.

\vskip 1cm
\noindent
{\bf 2. $\mathcal{N}=1$ superconformal mechanics with $osp(2|1)\oplus so(d)$ symmetry}
\vskip 0.5cm

The model of a free $\mathcal{N}=1$ superparticle of order $(2l+1)$ is described by the action functional
\bea\label{action}
S=\frac{1}{2}\int dt\,\lambda_{ij}(x_i x_j^{(2l+1)}-i\psi_i\psi_j^{(2l)}),
\eea
with $\lambda_{ij}$ in (\ref{bosonic_action}). As was shown in \cite{Masterov_1}, the action holds invariant under the $l > 1/2$ conformal Galilei group (no sum over $n$ below)
\bea\label{transf}
\begin{aligned}
&
L_{n}:\quad  \delta t=t^{n+1}a_n,\qquad\quad\; \delta x_i=l(n+1)t^n x_i a_n,\quad \delta\psi_i=(l-1/2)(n+1)t^n\psi_i a_n;
\\[2pt]
&
Q_{r}:\quad  \delta x_i=it^{r+1/2}\psi_i\alpha_r,\quad \delta\psi_i=(t^{r+1/2}\dot{x}_i-2l(r+1/2)x_i)\alpha_r;
\\[2pt]
&
C_i^{(n)}:\; \delta x_i = b_i^{(n)}t^{n}; \qquad\quad\; L_i^{(n)}:\, \delta\psi_i = \beta_i^{(n)}t^{n};
\qquad M_{ij}: \delta x_i=\omega_{ij}x_j,\quad \delta\psi_i=\omega_{ij}\psi_j,
\end{aligned}
\eea
where $a_n$,  $b_i^{(n)}$, $\alpha_r$, $\beta_i^{(n)}$ and $\omega_{ij} = -\omega_{ji}$ are infinitesimal parameters.

In order to construct a superfield formulation for (\ref{action}), the temporal coordinate $t$ is to be accompanied by a real Grassmann variable $\theta$ which give rise to a superfield
\bea
X_i(t,\theta)=x_i(t)+i\theta\psi_i(t)
\nonumber
\eea
and allow one to rewrite the action (\ref{action}) in the form
\bea\label{act1}
S=\frac{i^{2l+1}}{2}\int dt d\theta\,\lambda_{ij}X_i(t,\theta)\mathbb{D}^{4l+1}X_j(t,\theta),
\eea
where
\bea
\mathbb{D}=\frac{\vec{\partial}}{\partial\theta}-i\theta\frac{\partial}{\partial t}
\nonumber
\eea
is the covariant spinor derivative\footnote{Some technical details concerning the symmetry transformations of the action functional (\ref{act1}) are collected in Appendix.}.

Let us consider the superfield action of the form
\bea\label{general_action}
S=\frac{i^{2l+1}}{2}\int dt d\theta\left(\lambda_{ij}X_i(t,\theta)\mathbb{D}^{4l+1}X_j(t,\theta) + V\left(X_i,\mathbb{D}X_i,..,\mathbb{D}^{2l}X_i\right)\right),
\eea
where $V = V\left(X_i,\mathbb{D}X_i,..,\mathbb{D}^{2l}X_i\right)$ is a Grassmann-odd potential function.
The action functional is invariant under time translations as well as under supersymmetry transformations for any $V$.
The requirement that \eqref{general_action} be invariant under dilatations, special conformal transformations and superconformal transformations yields the constraints on $V$
\bea\label{equ_1}
&&
V + \sum_{n=0}^{2l-1}(2l-n)\mathbb{D}^{n}X_i\frac{\partial V}{\partial\mathbb{D}^{n}X_i} = 0,
\\[2pt]
&&\label{equ_2}
\sum_{n=0}^{l+\varepsilon - 1}(2l - n)\mathbb{D}^{2n}X_i\frac{\partial V}{\partial\mathbb{D}^{2n+1}X_i} - \sum_{n=1}^{l-\varepsilon}n\mathbb{D}^{2n-1}X_i\frac{\partial V}{\partial\mathbb{D}^{2n}X_i} = 0,
\\[2pt]
&&\label{equ_3}
\sum_{n=0}^{l-\varepsilon-1}(n+1)(2l-n)\mathbb{D}^{2n}X_i\frac{\partial V}{\partial\mathbb{D}^{2n+2}X_i} + \sum_{n=1}^{l+\varepsilon-1}n(2l-n)\mathbb{D}^{2n-1}X_i\frac{\partial V}{\partial\mathbb{D}^{2n+1}X_i} = 0,
\eea
with $\mathbb{D}^{0}X_i = X_i$. The parameter $\varepsilon$ is given by
\bea\label{vareps}
\varepsilon = \left\{
\begin{aligned}
&
\frac{1}{2}, && \mbox{ for half-integer }l,
\\[2pt]
&
0, && \mbox{ for integer }l.
\end{aligned}
\right.
\eea
The formal symbols $\sum\limits_{n=1}^{0}f(n)$ and $\sum\limits_{n=0}^{-1}f(n)$, which appear for $l=\frac{1}{2}$ and $l=1$ cases, are assumed to be equal to zero. Any solution of the equations (\ref{equ_1}) and (\ref{equ_2}) defines an $osp(2|1)$-invariant superconformal mechanics.

Notice that the equations (\ref{equ_2}) and (\ref{equ_3}) are not independent. It is straightforward to verify that they can be presented in the form
\bea
\mbox{(\ref{equ_2})}\;:\;\Omega V = 0,\qquad \mbox{(\ref{equ_3})}\;:\; -\Omega^2 V = 0,
\nonumber
\eea
where $\Omega$ is given by
\bea
\Omega = \sum_{n=0}^{l+\varepsilon -1}(2l - n)\mathbb{D}^{2n}X_i\frac{\partial}{\partial\mathbb{D}^{2n+1}X_i} - \sum_{n=1}^{l-\varepsilon}n\mathbb{D}^{2n-1}X_i\frac{\partial}{\partial\mathbb{D}^{2n}X_i}.
\nonumber
\eea

Let us analyze equation (\ref{equ_2}) in more detail. As the first step, we consider the instance $l=\frac{1}{2}$
\bea\label{equ_2_l=1/2}
X_i \frac{\partial V}{\partial\mathbb{D}X_i} = 0.
\eea
When $d=1$, the general solution of \eqref{equ_2_l=1/2} is an arbitrary function of the superfiled $X$, i.e.
\bea
V = V(X).
\nonumber
\eea
While for $d>1$, the general solution is constructed by the conventional method of characteristics
\bea\label{gs_l=1/2}
V = V\left(X_i,P_{ij}^{(1)}\right), \qquad P_{ij}^{(1)} = X_i\mathbb{D} X_j - X_j \mathbb{D}X_i.
\eea

As the next step, let us consider the equation (\ref{equ_2}) for the case of $l=1$
\bea
2X_i\frac{\partial V}{\partial\mathbb{D}X_i} - \mathbb{D}X_i\frac{\partial V}{\partial\mathbb{D}^2 X_i} = 0.
\nonumber
\eea
The analysis of the characteristic system for this equation yields the following solution
\bea\label{gs_l=1}
V = V(X_i,P_{ij}^{(1)},P_{ij}^{(2)}),
\eea
where we denoted
\bea
P_{ij}^{(2)} = X_i\mathbb{D}^2 X_j - X_j\mathbb{D}^2 X_i + \frac{1}{2}\mathbb{D}X_i\mathbb{D}X_j.
\nonumber
\eea

Let us obtain a solution to the equation (\ref{equ_2}) for an arbitrary value of the parameter $l$. Taking into account
the previously considered instances of $l=\frac{1}{2}$ and for $l=1$, it is natural to chose the following ansatz
\bea\label{ansatz}
\begin{aligned}
&
P_{ij}^{(2n-1)} = \sum_{k=0}^{n-1}\alpha_{k,n}\mathbb{D}^{2k}X_i\mathbb{D}^{2n-2k-1}X_j + \sum_{k=0}^{n-1}\beta_{k,n}\mathbb{D}^{2k+1}X_i\mathbb{D}^{2n-2k-2}X_j, && n=1,2,..,l+\varepsilon;
\\[2pt]
&
P_{ij}^{(2n)} = \sum_{k=0}^{n}\gamma_{k,n}\mathbb{D}^{2k}X_i\mathbb{D}^{2n-2k}X_j + \sum_{k=0}^{n - 1}\sigma_{k,n}\mathbb{D}^{2k+1}X_i\mathbb{D}^{2n-2k-1}X_j, && n=1,2,..,l-\varepsilon.
\end{aligned}
\eea
Substituting (\ref{ansatz}) into (\ref{equ_2}) one gets the recurrence relations for the coefficients $\alpha_{k,n}$, $\beta_{k,n}$, $\gamma_{k,n}$, and $\sigma_{k,n}$:
\begin{align}
&
(2l-n+k+1)\alpha_{k,n} + (2l-k)\beta_{k,n} = 0, && (k+1)\alpha_{k+1,n} - (n-k-1)\beta_{k,n} = 0,
\nonumber
\\[2pt]
&
(2l-k)\sigma_{k,n} - (n-k)\gamma_{k,n} = 0, && (2l - n + k + 1)\sigma_{k,n} + (k+1)\gamma_{k+1,n} = 0.
\nonumber
\end{align}
A solution to these equations reads
\bea\label{coeffs}
\begin{aligned}
&
\alpha_{k,n} = \frac{(-1)^{k}(2l-n+k)!(2l-k)!(n-1)!}{(2l-n)!(2l)!(n-k-1)!k!}, && \beta_{k,n} = -\frac{2l-n+k+1}{2l-k}\alpha_{k,n},
\\[2pt]
&
\gamma_{k,n} = \frac{(-1)^{k}(2l-n+k)!(2l-k)!n!}{(2l-n)!(2l)!(n-k)!k!}, && \sigma_{k,n} = \frac{n-k}{2l-k}\gamma_{k,n}.
\end{aligned}
\eea

A few comments are in order. Firstly, by construction, the polynomials $P_{ij}^{(k)}$ are Grassmann-odd functions for odd $k$ and Grassmann-even functions otherwise. Secondly, the coefficients (\ref{coeffs}) possess the following properties
\bea
\begin{aligned}
&
\alpha_{n-k-1,n} = (-1)^{n}\beta_{k,n}, &&\qquad && \beta_{n-k-1,n} = (-1)^{n}\alpha_{k,n},
\\[2pt]
&
\gamma_{n-k,n} = (-1)^{n}\gamma_{k,n}, && && \sigma_{n-k-1,n} = (-1)^{n-1}\sigma_{k,n}.
\end{aligned}
\nonumber
\eea
Due to these properties, the polynomials $P_{ij}^{(2n-1)}$ and $P_{ij}^{(2n)}$ are antisymmetric for odd $n$ and symmetric for even $n$, i.e.
\bea
&&
P_{ij}^{(2n-1)} = (-1)^{n} P_{ji}^{(2n-1)},\qquad P_{ij}^{(2n)} = (-1)^{n}P_{ji}^{(2n)}.
\nonumber
\eea

Thirdly, abbreviating $P_{ij}^{(0)} = X_i X_j$, a solution to equation \eqref{equ_2} can be written as follows
\bea
V = V\left(P_{ij}^{(0)},P_{ij}^{(1)},..,P_{ij}^{(2l)}\right).
\nonumber
\eea
In terms of the variables $P_{ij}^{(n)}$, the equation (\ref{equ_1}) takes the form
\bea\label{potential_equ}
V + \sum_{n=0}^{2l}(4l-n)P_{ij}^{(n)}\frac{\partial V}{\partial P_{ij}^{(n)}} = 0.
\eea
Any solution of this equation defines an $\mathcal{N}=1$ superconformal mechanics with $osp(2|1)\oplus so(d)$ symmetry provided $V$ transforms as a scalar under space rotations. In the next section we will consider a few examples of such superconformal mechanics.

\vskip 1cm
\noindent
{\bf 3. Examples}
\vskip 0.5cm

\noindent
{\it 3.1. The second-order $osp(2|1)\oplus so(d)$ invariant superconformal mechanics $\left(l=\frac{1}{2}\right)$}
\vskip 0.5cm

Let us consider a way to obtain $so(d)$-invariant solutions of the equation \eqref{potential_equ} for the case of $l=\frac{1}{2}$. In accord with \eqref{gs_l=1/2}, a rotationally invariant potential function $V$ may depend
on any scalars which can be constructed from the polynomials
\bea\label{polynoms_l=1/2}
&&
P_{ij}^{(0)} = X_i X_j, \qquad P_{ij}^{(1)} = X_i \mathbb{D}X_j - X_j \mathbb{D}X_i.
\eea
Taking into account that the function $P_{ij}^{(1)}$ is Grassmann-odd and antisymmetric, one may construct scalars by making use of $P_{ij}^{(0)}$ only. The simplest rotationally invariant combination is
\bea
\mathbb{P}_{0} = P_{ii}^{(0)} = X_i X_i.
\nonumber
\eea

In agreement with Eq. (\ref{potential_equ}), the potential function $V = V(\mathbb{P}_{0})$ satisfies the equation
\bea
V + 2\mathbb{P}_0 \frac{dV}{d\mathbb{P}_0} = 0.
\nonumber
\eea
The solution of this equation has the form
\bea\label{bad_function}
V = \frac{g}{\sqrt{\mathbb{P}_0}},
\eea
where $g$ is a coupling constant. By construction, the function $V$ must be Grassmann-odd. But this condition can be met only if $g$ is Grassmann-odd\footnote{See also related discussion in \cite{Rodrigues}.}. On physical grounds we discard such potential functions from our consideration.

To obtain a more appropriate rotationally invariant solution to Eq. \eqref{potential_equ}, we need to have Grassmann-odd scalars which can be constructed from polynomials \eqref{polynoms_l=1/2}. For $d=2$, a scalar of such a type can be obtained by contracting the Levi-Civit\'{a} symbol with $P_{ij}^{(1)}$, i.e.
\bea
\mathbb{P}_{1} = \epsilon_{ij}P_{ij}^{(1)} = 2P_{12}^{(1)}.
\nonumber
\eea
The equation (\ref{potential_equ}) for the function $V = V\left(\mathbb{P}_{0},\mathbb{P}_{1}\right)$ takes the form
\bea
V + 2\mathbb{P}_{0}\frac{\partial V}{\partial\mathbb{P}_{0}} + \mathbb{P}_1\frac{\partial V}{\partial\mathbb{P}_1} = 0,
\nonumber
\eea
whose general solution reads
\bea\label{potential_l=1/2}
V  = \frac{g}{\sqrt{\mathbb{P}_{0}}} + \frac{i\gamma\mathbb{P}_{1}}{\mathbb{P}_0}.
\eea
The first term reproduces the inappropriate potential function \eqref{bad_function} revealed above,
while the requirement for the second term in \eqref{potential_l=1/2} to be Grassmann-odd is satisfied provided the coupling constant $\gamma$ is Grassmann-even.

Let us consider the superfield action
\bea\label{SUSY_l=1/2}
S=-\frac{1}{2}\int dt d\theta\left(X_i(t,\theta)\mathbb{D}^{3}X_i(t,\theta) + \frac{i\gamma\mathbb{P}_{1}}{\mathbb{P}_0}\right).
\eea
Being rewritten in components, it takes the form
\bea
&&
S = \frac{1}{2} \int dt \left(x_i\ddot{x}_i - i\psi_i\dot{\psi}_{i} - \frac{2\gamma\epsilon_{ij}x_i\dot{x}_j}{x_k x_k}\right)
= \frac{1}{2} \int dt \left(x_i\ddot{x}_i - i\psi_i\dot{\psi}_{i} - 2\gamma\frac{d}{dt}\left(\arctan{\frac{x_2}{x_1}}\right)\right).
\nonumber
\eea
Thus, the model (\ref{SUSY_l=1/2}) corresponds to a free $\mathcal{N}=1$ second-order superparticle.

\vskip 0.5cm

\noindent
{\it 3.2. The third-order $osp(2|1)\oplus so(2)$ invariant superconformal mechanics $\left(l=1\right)$}
\vskip 0.5cm

When constructing $osp(2|1)\oplus so(2)$-invariant potentials for the third-order $\mathcal{N}=1$ superconformal mechanics, we have at our disposal one more polynomial
\bea
P_{ij}^{(2)} = X_i\mathbb{D}^{2}X_j - X_j\mathbb{D}^{2}X_i + \frac{1}{2}\mathbb{D}X_i\mathbb{D}X_j,
\nonumber
\eea
which is Grassmann-even and antisymmetric. As a consequence, the number of possible rotationally invariant combinations increases. For example, one can use
\bea\label{scalars_l=1}
\begin{aligned}
&
1) && \mathbb{P}_{0} = P^{(0)}_{ii} && \mbox{ - even},
\\[2pt]
&
2) && \mathbb{P}_1 = \epsilon_{ij}P_{ij}^{(1)} && \mbox{ - odd},
\\[2pt]
&
3) && \mathbb{P}_2 = \epsilon_{ij}P_{ij}^{(2)} && \mbox{ - even},
\\[2pt]
&
4) && \mathbb{P}_{12} = P^{(1)}_{ij} P^{(2)}_{ij} && \mbox{ - odd},
\\[2pt]
&
5) && \mathbb{P}_{22} = P^{(2)}_{ij} P^{(2)}_{ij} && \mbox{ - even},
\end{aligned}
\eea
as independent variables on which the potential function $V$ depends. In this case the equation (\ref{potential_equ}) can be rewritten as
\bea
V + 4\mathbb{P}_{0}\frac{\partial V}{\partial\mathbb{P}_{0}} + 3\mathbb{P}_{1}\frac{\partial V}{\partial\mathbb{P}_1} + 2\mathbb{P}_{2}\frac{\partial V}{\partial\mathbb{P}_2} + 5\mathbb{P}_{12}\frac{\partial V}{\partial\mathbb{P}_{12}} + 4\mathbb{P}_{22}\frac{\partial V}{\partial\mathbb{P}_{22}} = 0.
\nonumber
\eea
The general solution of this equation reads
\bea\label{potential_l=1}
V = \frac{1}{\sqrt[4]{\mathbb{P}_{0}}}\;\Lambda\left(\frac{\mathbb{P}_1}{\sqrt[4]{\mathbb{P}_{0}}^3}, \frac{\mathbb{P}_2}{\sqrt{\mathbb{P}_{0}}}, \frac{\mathbb{P}_{12}}{\sqrt[4]{\mathbb{P}_{0}}^5}, \frac{\mathbb{P}_{22}}{\mathbb{P}_{0}}\right),
\eea
where $\Lambda$ is an arbitrary function. The Taylor expansion of the function (\ref{potential_l=1}) in the Grassmann-odd variables can be divided into two parts
\bea
&
V = U + W,
\nonumber
\eea
where
\begin{align}
&\label{V_F}
U = \frac{1}{\sqrt[4]{\mathbb{P}_0}}\cdot\Psi_1\left(\frac{\mathbb{P}_2}{\sqrt{\mathbb{P}_{0}}},\frac{\mathbb{P}_{22}}{\mathbb{P}_{0}}\right) + \frac{\mathbb{P}_{1}\mathbb{P}_{12}}{\sqrt[4]{\mathbb{P}_{0}}^9}\cdot\Psi_2\left(\frac{\mathbb{P}_2}{\sqrt{\mathbb{P}_{0}}},\frac{\mathbb{P}_{22}}{\mathbb{P}_{0}}\right),
\\[2pt]
&\label{V_B}
W = \frac{\mathbb{P}_1}{\mathbb{P}_0}\cdot\Phi_1\left(\frac{\mathbb{P}_2}{\sqrt{\mathbb{P}_{0}}},\frac{\mathbb{P}_{22}}{\mathbb{P}_{0}}\right) + \frac{\mathbb{P}_{12}}{\sqrt{\mathbb{P}_{0}}^3}\cdot\Phi_2\left(\frac{\mathbb{P}_2}{\sqrt{\mathbb{P}_{0}}},\frac{\mathbb{P}_{22}}{\mathbb{P}_{0}}\right).
\end{align}
Here $\Psi_1$, $\Psi_2$, $\Phi_1$, and $\Phi_2$ are arbitrary functions.

Any potential in (\ref{V_F}) is Grassmann-odd only if the corresponding coupling constant is also Grassmann-odd. At the same time the potential functions in (\ref{V_B}) are more appropriate for our consideration. As an example, let us set $\Phi_1 = 0$, $\Phi_2=-i\gamma/2$ in \eqref{V_B} and consider the model which is described by the superfield action functional
\bea
S = -\frac{i}{2} \int dt d\theta \left(\epsilon_{ij}X_i\mathbb{D}^5 X_j - \frac{i\gamma\mathbb{P}_{12}}{2\sqrt{\mathbb{P}_{0}}^3}\right)
\nonumber
\eea
which has the component form
\bea\label{model_l=1}
\begin{aligned}
&
S = \frac{1}{2}\int dt \left(\epsilon_{ij}x_i\dddot{x}_j - i\epsilon_{ij}\psi_i\ddot{\psi}_j + \gamma\frac{(\epsilon_{ij}x_i\dot{x}_j)^2 + i(\epsilon_{ij}x_i\psi_j)(\epsilon_{kp}x_k\dot{\psi}_p)}{\sqrt{(x_s x_s)^3}}\right).
\end{aligned}
\eea
The invariance of this action under the $osp(2|1)\oplus so(2)$ transformations from (\ref{transf}) yields the following conserved charges
\bea
&&
\mathcal{L}_{-1} = \epsilon_{ij}\dot{x}_i\ddot{x}_j - \frac{i}{2}\epsilon_{ij}\dot{\psi}_i\dot{\psi}_j - \frac{\gamma}{2}\frac{(\epsilon_{ij}x_i\dot{x}_j)^2}{\sqrt{(x_s x_s)^3}},
\nonumber
\\[2pt]
&&
\mathcal{L}_{0} = t \mathcal{L}_{-1} - \epsilon_{ij}x_i\ddot{x}_j + \frac{i}{2}\epsilon_{ij}\psi_i\dot{\psi}_j,
\nonumber
\\[2pt]
&&
\mathcal{L}_{1} = -t^2 \mathcal{L}_{-1} + 2t \mathcal{L}_{0} + 2\epsilon_{ij}x_i\dot{x}_j - \frac{i}{2}\epsilon_{ij}\psi_i\psi_j,
\nonumber\\[2pt]
&&
\mathcal{Q}_{-1/2} = i\epsilon_{ij}\psi_i\ddot{x}_j - i\epsilon_{ij}\dot{\psi}_i\dot{x}_j - \frac{i\gamma(\epsilon_{ij}x_i\psi_j)(\epsilon_{kp}x_k\dot{x}_p)}{\sqrt{(x_s x_s)^3}},
\nonumber
\\[2pt]
&&
\mathcal{Q}_{1/2} = t\mathcal{Q}_{-1/2} - i\epsilon_{ij}\psi_i\dot{x}_j + 2i\epsilon_{ij}\dot{\psi}_i x_j,
\nonumber
\\[2pt]
&&
\mathcal{M}_{12} = \frac{1}{2}\dot{x}_i\dot{x}_i - x_i\ddot{x}_i + i\psi_i\dot{\psi}_i - \frac{\gamma(x_i x_i)(\epsilon_{jk} x_j\dot{x}_k)}{\sqrt{(x_s x_s)^3}} + \frac{i\gamma}{2}\frac{(x_i\psi_i)(\epsilon_{jk}x_j\psi_k)}{\sqrt{(x_s x_s)^3}}.
\nonumber
\eea
Here and what follows we denote constants of the motion by the same letters which were used for designating the corresponding symmetry generators, but in a calligraphic style.
\vskip 0.5cm

\noindent
{\it 3.3. The fourth-order $osp(2|1)\oplus so(d)$ invariant superconformal mechanics $\left(l=\frac{3}{2}\right)$}
\vskip 0.5cm

As was mentioned above, in general, the superpotential for the fourth-order $osp(2|1)$-invariant superconformal mechanics is a function of $P_{ij}^{(n)}$ with $n=0,1,2,3$,
which obeys the equation \eqref{potential_equ}. The polynomials $P_{ij}^{(0)}$ and $P_{ij}^{(1)}$ are the same as in (\ref{polynoms_l=1/2}), but others are given by
\bea
&&
P_{ij}^{(2)} = X_i\mathbb{D}^2 X_j - X_j\mathbb{D}^{2} X_i +\frac{1}{3}\mathbb{D}X_i\mathbb{D}X_j,
\nonumber
\\[2pt]
&&
P_{ij}^{(3)} = X_i\mathbb{D}^{3}X_j + X_j \mathbb{D}^{3}X_i - \frac{2}{3}\left(\mathbb{D}X_i\mathbb{D}^{2}X_j + \mathbb{D}X_j\mathbb{D}^2 X_i\right).
\nonumber
\eea
Note that $P_{ij}^{(3)}$ is Grassmann-odd symmetric polynomial.

Taking into account the analysis in the preceding subsection, let us consider the superpotential which is a linear function of the anticommuting scalars $\mathbb{P}_{12}$ and $\mathbb{P}_{03} = P_{ij}^{(0)}P_{ij}^{(3)}$,
\bea\label{potential_l=3/2}
V = \mathbb{P}_{12} V_1\left(\mathbb{P}_{0},\mathbb{P}_{22}\right) + \mathbb{P}_{03} V_2\left(\mathbb{P}_{0},\mathbb{P}_{22}\right),
\eea
with $\mathbb{P}_{0}$, $\mathbb{P}_{12}$, and $\mathbb{P}_{22}$ defined in (\ref{scalars_l=1}). The ansatz \eqref{potential_l=3/2} solves \eqref{potential_equ} provided
\bea\label{solution_l=3/2}
&&
V_1 =  \frac{1}{\sqrt[3]{\mathbb{P}_{0}}^5}\Phi_1\left(\frac{\mathbb{P}_{22}}{\sqrt[3]{\mathbb{P}_{0}}^4}\right),\qquad
V_2 = \frac{1}{\sqrt[3]{\mathbb{P}_{0}}^5}\Phi_2\left(\frac{\mathbb{P}_{22}}{\sqrt[3]{\mathbb{P}_{0}}^4}\right),
\eea
where $\Phi_1$ and $\Phi_2$ are arbitrary functions.

As an example, let us consider the potential function
\bea\label{V_12}
V = -\frac{\gamma\mathbb{P}_{12}}{\sqrt[3]{\mathbb{P}_{0}}^5},
\eea
which corresponds to the component action 
\bea\label{model_l=3/2_1}
\;S = \frac{1}{2}\int dt  \left(x_i x_i^{(4)} - i \psi_i \psi_i^{(3)} + \gamma\frac{\dot{x}_i\dot{x}_i + i\psi_i\dot{\psi}_i}{\sqrt[3]{(x_k x_k)^2}} - \gamma\frac{(x_i\dot{x}_i)^2 + i(x_i \psi_i)(x_j\dot{\psi}_j) - \frac{i}{3}(x_i\psi_i)(\dot{x}_j\psi_j)}{\sqrt[3]{(x_k x_k)^{5}}}\right).
\nonumber
\eea
By conventional means one finds the integrals of motion
\bea\label{IM}
&&
\begin{aligned}
&
\mathcal{L}_{-1} = \dot{x}_i\dddot{x}_i -\frac{1}{2}\ddot{x}_i\ddot{x}_i -i\dot{\psi}_i\ddot{\psi}_i -
\frac{\gamma}{2}\left(\frac{\dot{x}_i\dot{x}_i}{\sqrt[3]{(x_j x_j)^2}} - \frac{(x_i\dot{x}_i)^2}{\sqrt[3]{(x_j x_j)^5}}\right),
\\[2pt]
&
\mathcal{L}_{0} = t\mathcal{L}_{-1} - \frac{3}{2}x_i\dddot{x}_i + \frac{1}{2}\dot{x}_i\ddot{x}_i +i\psi_i\ddot{\psi}_i,
\\[2pt]
&
\mathcal{L}_{1} = -t^2\mathcal{L}_{-1} + 2t\mathcal{L}_{0} -2\dot{x}_i\dot{x}_i + 3x_i\ddot{x},
\\[2pt]
&
\mathcal{Q}_{-1/2} = i\psi_i\dddot{x}_i - i\dot{\psi}_i\ddot{x}_i
+ i\ddot{\psi}_i\dot{x}_i - \frac{i\gamma\psi_i\dot{x}_i}{\sqrt[3]{(x_k x_k)^2}} + \frac{i\gamma(x_i\psi_i)(x_j\dot{x}_j)}{\sqrt[3]{(x_k x_k)^5}},
\\[2pt]
&
\mathcal{Q}_{1/2} = t\mathcal{Q}_{-1/2} -i\psi_i\ddot{x}_i + 2i\dot{\psi}_i\dot{x}_i - 3i\ddot{\psi}_i x_i,
\\[2pt]
&
\mathcal{M}_{ij} = x_{[i}\dddot{x}_{j]} - \dot{x}_{[i}\ddot{x}_{j]} - i\psi_{[i}\ddot{\psi}_{j]} + i\dot{\psi}_{i}\dot{\psi}_{j}
- \frac{\gamma x_{[i}\dot{x}_{j]}}{\sqrt[3]{(x_s x_s)^{2}}} + \frac{i\gamma\psi_{i}\psi_{j}}{\sqrt[3]{(x_s x_s)^2}} + \frac{2i\gamma}{3}\frac{x_{[i}\psi_{j]}(x_k\psi_k)}{\sqrt[3]{(x_s x_s)^5}},
\end{aligned}
\eea
associated with the $osp(2|1)\oplus so(d)$ symmetry.

Note that the superpotential \eqref{V_12} is proportional to $\mathbb{P}_{12} = P_{ij}^{(1)}P_{ij}^{(2)}$ and consequently it vanishes when $d=1$ because the polynomial $P_{ij}^{(1)}$ is antisymmetric.
Therefore, for one dimensional case, the conserved quantities \eqref{IM} go over to the corresponding expressions for a free $\mathcal{N}=1$ fourth-order superparticle. However, the potential functions \eqref{potential_l=3/2} with nonzero $V_2$  are viable in arbitrary dimension including $d=1$. As an example, one can consider 
\bea
V = \frac{\gamma\mathbb{P}_{03}}{2\sqrt[3]{\mathbb{P}_{0}}^5},
\nonumber
\eea
which corresponds to an $\mathcal{N}=1$ fourth-order superconformal mechanical system which is governed by the component action
\bea
S = \frac{1}{2}\int dt \left( x_i x_i^{(4)} - i \psi_i \psi_i^{(3)} + \gamma
\frac{\dot{x}_i\dot{x}_i + i\psi_i\dot{\psi}_i}{\sqrt[3]{(x_s x_s)^2}} -
\frac{2\gamma}{3}\frac{(x_i\dot{x}_i)^2 + i(x_i\psi_i)(x_j\dot{\psi}_j) - i(x_i\psi_i)(\dot{x}_j\psi_j)}{\sqrt[3]{(x_s x_s)^5}}\right).
\nonumber
\eea
Conserved quantities associated with the $osp(2|1)\oplus so(d)$ symmetry read
\bea
\begin{aligned}
&
\mathcal{L}_{-1} = \dot{x}_i\dddot{x}_i - \frac{1}{2}\ddot{x}_i\ddot{x}_i - i\dot{\psi}_i\ddot{\psi}_i - \frac{\gamma}{2}\frac{\dot{x}_i\dot{x}_i}{\sqrt[3]{(x_s x_s)^2}} + \frac{\gamma}{3}\frac{(x_i\dot{x}_i)^2}{\sqrt[3]{(x_s x_s)^5}},
\\[2pt]
&
\mathcal{L}_{0} = t\mathcal{L}_{-1} - \frac{3}{2}x_i\dddot{x}_i + \frac{1}{2}\dot{x}_i\ddot{x}_i + i \psi_i\ddot{\psi}_i + \frac{\gamma}{2}\frac{x_i\dot{x}_i}{\sqrt[3]{(x_s x_s)^2}},
\\[2pt]
&
\mathcal{L}_{1} = -t^2 \mathcal{L}_{-1} + 2t \mathcal{L}_{0} + 3x_i\ddot{x}_i - 2\dot{x}_i\dot{x}_i - 2i\psi_i\dot{\psi}_i - \frac{3\gamma}{2}\sqrt[3]{x_s x_s},
\\[2pt]
&
\mathcal{Q}_{-1/2} = i\psi_i\dddot{x}_i - i\dot{\psi}_i\ddot{x}_i + i\ddot{\psi}_i\dot{x}_i - \frac{i\gamma\psi_i\dot{x}_i}{\sqrt[3]{(x_k x_k)^2}} + \frac{2i\gamma}{3}\frac{(x_i\psi_i)(x_j\dot{x}_j)}{\sqrt[3]{(x_k x_k)^5}},
\\[2pt]
&
\mathcal{Q}_{1/2} = t\mathcal{Q}_{-1/2} -i\psi_i\ddot{x}_i + 2i\dot{\psi}_i\dot{x}_i - 3i\ddot{\psi}_i x_i + \frac{i\gamma x_i\psi_i}{\sqrt[3]{(x_s x_s)^2}},
\\[2pt]
&
\mathcal{M}_{ij} = x_{[i}\dddot{x}_{j]} - \dot{x}_{[i}\ddot{x}_{j]} - i\psi_{[i}\ddot{\psi}_{j]} + i\dot{\psi}_{i}\dot{\psi}_{j}
- \frac{\gamma x_{[i}\dot{x}_{j]}}{\sqrt[3]{(x_s x_s)^{2}}} + \frac{i\gamma\psi_{i}\psi_{j}}{\sqrt[3]{(x_s x_s)^2}} + \frac{2i\gamma}{3}\frac{x_{[i}\psi_{j]}(x_k\psi_k)}{\sqrt[3]{(x_s x_s)^5}}.
\end{aligned}
\nonumber
\eea

\vskip 1cm
\noindent
{\bf 4. Conclusion}
\vskip 0.5cm

To summarize, in this work we formulated the equations \eqref{equ_1}, \eqref{equ_2} which determine a potential function in an $\mathcal{N}=1$ higher derivative supersymmetric mechanics 
compatible with the $osp(2|1)\oplus so(d)$ symmetry and provided a few explicit examples. Our strategy to obtain $osp(2|1)\oplus so(d)$-invariant mechanics included the following steps:
\begin{itemize}
    \item list all the polynomials $P_{ij}^{(n)}$ in \eqref{ansatz} which correspond to the given order of a dynamical system;
    \item construct $so(d)$-invariant combinations from these polynomials;
    \item find $so(d)$-invariant solutions of the equation \eqref{potential_equ} in terms of the rotationally invariant combinations;
    \item restrict oneself to a subset of the solutions with Grassmann-even coupling constants.
\end{itemize}

Turning to further possible developments, it would be interesting to construct quantum mechanical counterparts of the models described above. In this context, the method of conformal automorphisms previously developed in Ref. \cite{autom} might be a reasonable starting point. A generalization of the present analysis to cover interaction potentials which preserve the full $\mathcal{N}=1$ $l$-conformal Galilei supersymmetry is also worth studying.

Classical stability of higher-derivartive systems is an important issue. Recently the concept of the so-called Lagrange anchor \cite{KLS_1} was successfully applied to construct and investigate stable higher derivative mechanical systems (see, e.g., \cite{KLS_2}-\cite{KLS_4}). It is interesting to see how whether that approach can be adopted to the case of higher derivative $\mathcal{N}=1$ superconformal mechanics.

In general, $l$-conformal Galilei algebra can be realized in nonrelativistic space-time with universal cosmological
attraction or repulsion by means of Niederer's transformations. It would be interesting to obtain a superfield analogue of such transformation as well as to construct Newton-Hooke counterparts of the $\mathcal{N}=1$ superconformal mechanics described above.

\vskip 0.5cm
\noindent
{\bf Acknowlegments}
\vskip 0.5cm

We thank A. Galajinsky for useful comments on the manuscript. This work was supported by the Russian Science Foundation, grant No 19-11-00005.

\vskip 0.5cm
\noindent
{\bf Appendix. The symmetry transformations for the action functional (\ref{act1})}
\vskip 0.5cm

It is straightforward to verify that the differential operators
\bea\label{gen}
\begin{aligned}
&
L_{n}=t^{n+1}\frac{\partial}{\partial t}+\frac{1}{2}(n+1)t^n\theta\frac{\vec{\partial}}{\partial\theta} + l(n+1)t^n X_i\frac{\partial}{\partial X_i},
\\[2pt]
&
Q_{r}=i\theta t^{r+1/2}\frac{\partial}{\partial t} + t^{r+1/2}\frac{\vec{\partial}}{\partial\theta}+2l(r+1/2)\theta X_i\frac{\partial}{\partial X_i},
\\[2pt]
&
C_i^{(n)}=t^n\frac{\partial}{\partial X_i},\qquad L_i^{(n)}=i\theta t^n\frac{\partial}{\partial X_i},\qquad M_{ij}=X_i\frac{\partial}{\partial X_j}-X_j\frac{\partial}{\partial X_i},
\end{aligned}
\eea
obey the (anti)commutation relations of the $\mathcal{N}=1$ $l$-conformal Galilei superalgebra
\bea\label{lCG}
\begin{aligned}
&
[L_n,L_m]=(m-n)L_{m+n}, && [L_n,C_i^{(m)}]=(m-l(n+1))C_i^{(n+m)},
\\[2pt]
&
\{Q_{r},Q_s\}=2iL_{r+s}, && [L_n,L_i^{(m)}]=(m-(l-1/2)(p+1))L_i^{(n+m)},
\\[2pt]
&
[L_n,Q_r]=(r-n/2)Q_{n+r}, && [Q_r,C_i^{(n)}]=(n-2l(r+1/2))L_i^{(r+n-1/2)},
\\[2pt]
&
\{Q_r,L_i^{(n)}\}=iC_i^{(n+r+1/2)}, && [M_{ij},L_k^{(n)}]=\delta_{ik}L_j^{(n)}-\delta_{jk}L_i^{(n)},
\\[2pt]
&
[M_{ij},C_k^{(n)}]=\delta_{ik}C_j^{(n)}-\delta_{jk}C_i^{(n)}, && [M_{ij},M_{kl}]=\delta_{ik}M_{jl}+\delta_{jl}M_{ik}-\delta_{il}M_{jk}-\delta_{jk}M_{il}.
\end{aligned}
\eea
and generate the symmetry transformations for the action functional (\ref{act1}). As an example, let us demonstrate how one obtains the superconformal transformations by making use of $Q_{\frac{1}{2}}$ in (\ref{gen}). The symmetry transformations, which correspond to this generator, read
\bea\label{expr_1}
t'=t+it\alpha_{\frac{1}{2}}\theta,\quad \theta'=\theta+t\alpha_{\frac{1}{2}},\quad {X'}_i(t',\theta')=X_i(t,\theta)+2il\alpha_{\frac{1}{2}}\theta X_i(t,\theta).
\eea
On the other hand, we have
\bea\label{expr_2}
{X'}_i(t',\theta')={X'}_i(t,\theta)+it\alpha_{\frac{1}{2}}\theta\frac{\partial X_i}{\partial t}+it\alpha_{\frac{1}{2}}\frac{\vec{\partial}X_i}{\partial\theta}.
\eea
By comparing \eqref{expr_1} and \eqref{expr_2}, one obtains
\bea
\delta^* X_i(t,\theta) &\equiv& {X'}_i(t,\theta)-X_i(t,\theta) = it\psi_i\alpha_{\frac{1}{2}}+i\theta (t\dot{x}_i-2lx_i)\alpha_{\frac{1}{2}}.
\nonumber
\eea
By taking into account that $\delta^* X_i(t,\theta) = \delta x_i+i\theta\delta\psi_i$, we reproduce the superconformal transformations in (\ref{transf}).

If the action functional (\ref{act1}) holds invariant under the transformations
\bea
t'=t+\delta t,\qquad \theta'=\theta+\delta\theta, \qquad {X'}_i(t',\theta')=X_i(t,\theta)+\delta X_i(t,\theta)
\nonumber
\eea
up to a partial time derivative of some function $F=F(t,\theta)$, i.e. $\delta S=\int dt d\theta\,\frac{\partial F(t,\theta)}{\partial t}$, then the corresponding conserved quantity can be derived from the expression\footnote{Here we use the notations $\partial_t X_i=\frac{\partial X_i(t,\theta)}{\partial t}$, $\vec{\partial}_{\theta}X_i=\frac{\vec{\partial}X_i(t,\theta)}{\partial\theta}$. $L$ is the Lagrangian of the model (\ref{act1}).}
\bea
K = \frac{\vec{\partial}}{\partial\theta}\left[L\delta t+\sum_{n=0}^{2l}\frac{\partial^{n}}{\partial t^n}\left(\delta X_i-\delta t \partial_t X_i-\delta\theta\vec{\partial}_{\theta}X_i\right)\sum_{k=0}^{2l-n}(-1)^k\frac{\partial^k}{\partial t^k}\frac{\partial L}{\partial(\partial_t^{n+k+1}X_i)}-F\right]
\nonumber
\eea
by discarding parameter of the transformation. Conserved charges obtained in such a way are the superfield analogues of the integrals of motion associated with the symmetries (\ref{transf}) of the component action (\ref{action}).

\fontsize{10}{13}\selectfont

\end{document}